\documentclass[12pt]{iopart}
\usepackage{iopams}

\begin{document}

\title[Partition functions and symmetric polynomials]
{Partition functions and symmetric polynomials}

\author{Heinz-J\"urgen Schmidt\dag
\footnote[3]{To
whom correspondence should be addressed (hschmidt@uos.de)}
and J\"urgen Schnack\dag}

\address{\dag\ Universit\"at Osnabr\"uck, Fachbereich Physik,
Barbarastr. 7, 49069 Osnabr\"uck, Germany}

\begin{abstract}
We find a close correspondence between certain partition functions
of ideal quantum gases and certain symmetric polynomials. Due to
this correspondence it can be shown that a number of
thermodynamic identities which have recently been considered are
essentially of combinatorical origin and known for a long time
as theorems on symmetric polynomials. For example, a recurrence
relation for partition functions appearing in the textbook of
P.~Landsberg is nothing else but Newton's identity in disguised
form.  Conversely, a certain theorem on symmetric polynomials
translates into a new and unexpected relation between fermionic
and bosonic partition functions, which can be used to express
the former by means of the latter and vice versa.
\end{abstract}



\maketitle

\section{Introduction}

The theory of ideal quantum gases, although a standard section
in every modern textbook on statistical mechanics, is far from
being a closed field of research. On one hand, there are
continuous attempts towards a deeper understanding of the
foundations of the field and to disclose unexpected regularities
and relations between known facts. As examples we quote the work
of M.~H.~Lee \cite{Lee:JMP95,Lee:PRE97A,Lee:PRE97B} on a unified
treatment of Fermi and Bose gases and our earlier paper on a
fermion-boson symmetry for harmonic oscillators of odd space
dimensions \cite{ScS:PA99}. Also this article is intended to
contribute to such foundational questions

On the other hand, there are various applications of the theory
of ideal gases in almost all areas of modern physics, despite
the apparent over-simplification of non-interacting
particles. In nuclear physics the shell model is a prominent
representative, which at low excitation energies describes the
motion of noninteracting fermions in a common (mean) field
\cite{RiS80}.  In solid state physics of one space dimension
interacting fermions may be described as noninteracting bose
gases (Luttinger liquids) \cite{Tom:PTP50,Lut:JMP63,ScM:AJP96}.
The crucial point is that the specific heat of $N$ fermions and
bosons is the same for equidistant energy spectra, for recents
works see \cite{ScS:PA98,CL:PRA01}.

In atomic physics the concept of ideal quantum gases helped to
understand experiments on magnetically trapped atomic vapours
\cite{AEM:S95,DMA:PRL95,KeD:PRA96,BSH:PRL97}. Here the system
can be well described as an ideal quantum gas contained in an
external harmonic oscillator potential, for an overview see
\cite{DGP:RMP99}. Also the case of constant and small particle
number $N$ turns out to be of interest, thereby excluding the
use of a grand canonical ensemble ($N$ fluctuating) and/or the
thermodynamic limit ($N\rightarrow\infty$). Examples of
applications, except the above-mentioned trapped systems, are
quantum dots \cite{Alh:RMP00} and fluorescence from a few electrons
\cite{ACP:PRB00}. For a comparison
of canonical and grand canonical results in a two-state
system see also \cite{LH:JSP88}.

In the context of these small quantum systems it is sometimes of
interest to explicitely calculate the $N-$particle partition
function $Z_N^\pm$, where $+$ stands for bosons and $-$ for
fermions.  For this purpose a certain recurrence relation for
$Z_N^\pm$ has recently been used by several authors
\cite{BoF:JCP93,BLD:PRE97A,ScS:PA98,ScS:PA99,ACP:PRB00}.  It
seems that none of the authors, including us, was aware that
this recurrence relation could be traced back to the
time-honoured textbook of P.~Landsberg \cite{Lan61}. Even more
remarkable we find  that, in some sense,
Landsberg's formula was already discovered by Isaac Newton. Of
course, Newton did not know anything about partition functions
and Fermi/Bose statistics.  However, there is an intimate
connection between partition functions and symmetric
polynomials, which allows to translate certain theorems on
symmetric functions into statements about physical partition
functions and vice versa. This will be the main subject of the
present article.

The basic idea will be spelled out in section 2.
Since we write this article for people who may be experts in
either field, combinatorics and statistical mechanics, but not
in both, we will give some rather elementary introductions into
the field of symmetric polynomials (section 3) and partition
functions (section 4). Section 4 also contains the ``dictionary"
needed for the translation between statements of the respective
fields. Thus we obtain the following results, which now appear
obvious, but have not yet been explicitely mentioned in the physics
literature, namely that $Z_N^+$ may be expressed by the $Z_n^-,
n\le N$ and vice versa. The most compact expression of this is
the statement that the grand canonical partition
functions satisfy ${\cal Z}^+(z) {\cal Z}^-(-z) =1 $. In section
5 we add some closing remarks.

\section{The basic idea}
Before going into details we will describe the basic idea
of the correspondence between
partition functions and symmetric polynomials and illustrate this idea for the
simple example of a system with $L=3$ energy levels and $N=2$ particles.\\

It is well-known that the eigenstates of the $N-$particle Hamiltonian
$H_N^\pm$ ($+$ for bosons, $-$ for fermions) of $N$ non-interacting particles
can be labelled by occupation number sequences $(n_1,n_2,\ldots,n_L)$,
where $\sum_{\ell=1}^L n_\ell =N$ and $n_\ell=0,1$ for fermions. The latter
condition expresses the Pauli principle. Equivalently, the eigenstates are in
$1~:~1$-correspondence to the \underline{monomials}
$x_1^{n_1} x_2^{n_2}\ldots x_L^{n_L}$ of degree $N$, where the
$x_1,x_2,\ldots,x_L$  are (symbolic) commuting variables. In our example,
we obtain the monomials $x_1 x_2, x_1 x_3, x_2 x_3$ for the fermionic case,
and, additionally, $x_1^2, x_2^2, x_3^2$, for the bosonic case. \\

The eigenvalue of $H_N^\pm$ corresponding to the eigenstate labelled by
$x_1^{n_1} x_2^{n_2}\ldots x_L^{n_L}$ is given by $\sum_{\ell=1}^L E_\ell n_\ell$,
if $E_\ell$ denotes the $\ell$-th energy eigenvalue of the one-particle-Hamiltonian.
Hence the corresponding eigenvalue of $\exp(-\beta H_N^\pm)$, where, as usual,
$\beta=\frac{1}{k_B T}$ denotes the inverse temperature, will be
\begin{equation}\label{}
\exp(-\beta \sum_{\ell=1}^L E_\ell n_\ell)= (e^{-\beta E_1})^{n_1}\ldots
(e^{-\beta E_L})^{n_L}.
\end{equation}
This eigenvalue is nothing else but the value of the monomial
$x_1^{n_1} x_2^{n_2}\ldots x_L^{n_L}$ evaluated at $x_1=e^{-\beta E_1},\ldots,
x_L=e^{-\beta E_L}$. To obtain the trace $\mbox{Tr}\exp(-\beta H_N^\pm)$
we have to sum over all eigenstates, or, equivalently, over all monomials
of degree $N$
(subject to the constraint $n_\ell=0,1$ in the fermionic case). Interchanging
summation and evaluation, we may first consider the bosonic polynomial $b_N$
(resp.~the fermionic $f_N$)
obtained by summing up all monomials of degree $N$ (satisfying the above constraint
in the case of $f_N$) and then, in a second step, evaluate this polynomial
at $x_\ell=e^{-\beta E_\ell}, \ell=1,\ldots,L$ in order to obtain
$\mbox{Tr}\exp(-\beta H_N^\pm)$. The latter is usually called the
\underline{partition function} of the system. In our example,
$b_2=x_1^2+x_2^2+x_3^2+x_1x_2+x_1x_3+x_2x_3$ and
$f_2=x_1x_2+x_1x_3+x_2x_3$. Both polynomials are symmetric w.~r.~t.~any
permutation of their arguments $x_1, x_2, x_3$. This also holds
in the general case.\\

Thus we have established a connection between systems of $N$ non-interacting
bosons (resp.~fermions) and certain symmetric polynomials $b_N$ (resp.~$f_N$).
Specification of the one-particle Hamiltonian then corresponds to an evaluation
of these polynomials at $x_\ell=e^{-\beta E_\ell}, \ell=1,\ldots,L$ and yields
the particular partition function.

\section{Symmetrical polynomials}

We collect some well-known definitions and results on symmetric
polynomials which are relevant for our purposes. Throughout we
consider polynomials in a finite number of (symbolic) variables
$x_1,\ldots,x_n$.  Most results also hold for the case of
infinitely many variables ($n=\infty$) which is of physical
interest, where the polynomials become formal power series. We
will only indicate those cases where $n=\infty$ becomes
problematic. \\ A polynomial $p$ is called \underline{symmetric}
iff $p(x_1,\ldots,x_n)=p(x_{\sigma(1)},\ldots,x_{\sigma(n)})$
for all permutations $\sigma\in {\cal S}_n$. The
\underline{elementary symmetric} polynomials $f_m$ can be
defined through their generating function
\begin{equation}\label{1}
\prod_{\ell =1}^n (x+x_\ell) = \sum_{m=0}^n   f_m x^{n-m}, \quad m=0,\ldots,n.
\end{equation}
For example,
\begin{eqnarray}\label{2}
f_0 & = & 1,\\
f_1 & = & x_1+x_2 +\ldots +x_n,\\
f_2 & = & \sum_{i<j}x_i x_j,\\  \nonumber
\vdots & & \\
f_n &  =&  x_1 x_2 \cdots x_n.
\end{eqnarray}
The letter ``f" for elementary symmetric polynomials is unusual,
but chosen here to stress the association to ``fermions". One
important property of the $f$'s is that $(f_1,\ldots,f_n)$ forms
a \underline{basis} in the sense that any symmetric polynomial
can be uniquely written as a polynomial of the $f_m$.  For
example, if $n=2$,
\begin{equation}\label{3}
x_1^2 + x_2^2 =(x_1 + x_2)^2 -2 x_1 x_2 = f_1^2 - 2 f_2.
\end{equation}
Of course, there exist other bases of symmetric polynomials. For
our purposes, the following ones are the most important:

Let $b_\ell$ denote the sum of all monomials of degree $\ell$,
for example
\begin{eqnarray}\label{4}
b_0 & = & 1,\\
b_1 & = & x_1+x_2 +\ldots +x_n,\\
b_2 & = & \sum_{i\le j}x_i x_j,\\
\vdots & &   \nonumber
\end{eqnarray}

Obviously the $b_\ell$ are symmetric polynomials, sometimes called
\underline{complete} \underline{symmetric} polynomials
(our ``b" refers to ``bosons").
Also the sequence $(b_0,\ldots,b_n)$ forms a basis in the ring of all symmetric
polynomials in $n$ variables, as we will see below. For example, if $n=2$,
\begin{equation}\label{5}
x_1^2+x_2^2 = 2(x_1^2+x_2^2+x_2 x_2)-(x_1+x_2)^2 = 2 b_2-b_1^2.
\end{equation}
Another basis is given by the \underline{power sums}
\begin{equation}\label{6}
s_\ell = \sum_{i=1}^n x_i^\ell,\quad \ell=0,\ldots, n.
\end{equation}
($s_0=n$ has to be redefined in the case $n=\infty$.) In order
to prove that $(s_0,\ldots,s_n)$ forms a basis, it would suffice
to show that every elementary symmetric polynomial $f_m$ could
be expressed as a polynomial of the $s_\ell$.  This, in turn,
follows immediately from the \underline{Newton identities}, as
they are called in Ref.~\cite{CLS:1992},
\begin{equation}\label{7}
m f_m = \sum_{k=0}^{m-1} (-1)^{k+1} f_k s_{m-k}  ,\quad 1\le m \le n.
\end{equation}
These identities could be viewed as recursion relations for the
$f_m$.  Iterative substitution then yields the $f_m$ solely in
terms of the $s_\ell$, for example
\begin{equation}\label{8}
f_3=\frac{1}{6} (s_1^3-3 s_1 s_2 +2 s_3).
\end{equation}
Explicit formulae for these representations will be given
below.

Now consider the case of the $b_m$. Here we have analogous
identities which immediately imply that $(b_0,\ldots,b_n)$ will
form a basis:
\begin{equation}\label{9}
\sum_{k=0}^m (-1)^k f_k b_{m-k} =0, \quad 1\le m \le n.
\end{equation}
These identities can most compactly be written in terms of the
\underline{generating functions}
\begin{eqnarray}\label{10}
F(z) & =\sum_{r=0}^\infty f_r z^r &= \prod_{\ell=1}^\infty (1+x_\ell z),\\
B(z) & =\sum_{r=0}^\infty b_r z^r &= \prod_{\ell=1}^\infty (1-x_\ell z)^{-1} \label{10a}
\end{eqnarray}
as
\begin{equation}\label{10b}
B(z)F(-z)=1.
\end{equation}

Finally we give the explicit representations of the $f_m$ and
$b_m$ in terms of the power sums $s_\ell$, as well as the
representations of the $f_m$ in terms of the $b_k$ and vice
versa. To write these representations in compact form, it is
convenient to use \underline{partitions}. A partition $\lambda$
of a positive integer $N$, written as $\lambda\vdash N$, is a
way to write $N$ as a sum, irrespective of order, in other
words, a non-increasing sequence of positive integers
$(\lambda_1,\lambda_2,\ldots,\lambda_p)$ such that $\sum_{i=1}^p
\lambda_i = N$. $\ell(\lambda)\equiv p$ is called the
\underline{length} of a partition. Another notation for
$\lambda$ is $(1^{m_1}2^{m_2}\ldots)$ where $m_i$ counts the
occurence of the integer $i$ in the partition
$(\lambda_1,\lambda_2,\ldots,\lambda_p)$.  Further let
$\zeta(\lambda)\equiv (\prod_{i=1}^p i^{m_i} m_i!)^{-1}$
and $\mu(m)$ denote the multinomial coefficient
\begin{equation}\label{10c}
\mu(m)\equiv \frac{(\sum_i m_i)!}{\prod_i m_i!}.
\end{equation}
Products of special symmetric polynomials introduced above
according to a partition $\lambda$ will be written with a
subscript $\lambda$, for example,
\begin{equation}\label{10d}
f_\lambda \equiv \prod_{\ell=1}^p f_{\lambda_\ell}.
\end{equation}
Then
\begin{equation}\label{11}
\left.
\begin{array}{l}
b_N \\ f_N
\end{array}
   \right\} =
\sum_{\lambda\vdash N} (\pm 1)^{N+\ell(\lambda)}\zeta(\lambda) s_\lambda,
\end{equation}
where the $+$sign refers to the $b$'s and the $-$sign  to the
$f$'s, and
\begin{equation}\label{12}
\left.
\begin{array}{l}
b_N \\ f_N
\end{array}
   \right\} =
\sum_{\lambda\vdash N} (-1)^{N+\ell(\lambda)}\mu(m)
\left\{
\begin{array}{l}
f_{\lambda} \\b_{\lambda}
\end{array}
   \right.   .
\end{equation}
Note the symmetry of the transformations $f \rightarrow b $ and
$b \rightarrow f $.  This is a consequence of the invariance of
the identities (\ref{9}) with respect to the interchange
$b\leftrightarrow f$ and of $b_0=f_0=1$.  Note that (\ref{12}),
according to (\ref{10a}), is essentially
equivalent to a well-known explicit expansion formula for
the reciprocal value of a power series. These and other explicit
relations between special symmetric polynomials are often also
written in terms of determinants, see \cite{Mac:1979}.

\section{Partition functions}
The canonical partition function plays a fundamental role in
statistical mechanics since most thermodynamic functions can be
derived from it. It is defined by
\begin{equation}\label{13}
Z(\beta)\equiv\Tr \exp(-\beta H),
\end{equation}
where $H$ denotes the Hamiltonian of the system. Sometimes one
writes $Z_N(\beta)$ in order to stress the dependance on the
number $N$ of particles in the system.  The grand canonical
partition function is defined by
\begin{equation}\label{13a}
{\cal Z}(z,\beta)\equiv\sum_{N=0}^\infty Z_N(\beta) z^N,
\end{equation}
where the variable $z$ is physically interpreted as the fugacity
of the system.

Here we will only consider ideal gases, i.~e.~the case where H
 is the sum of $N$ (identical) single-particle Hamiltonians
\begin{equation}\label{14}
H = \sum_{n=1}^N h_n,
\end{equation}
for example, $N$ fermions in a common potential or mean field.

Let $E_\ell$ denote the energy eigenvalues of $h_n$, counted in
such a way that several $E_\ell$ have the same value in case of
degeneracy. Then the bosonic($+$) or fermionic($-$) partition
functions depend only on the $E_\ell$ and can be written as
\begin{equation}\label{15a}
Z_N^-(\beta) = \sum_{i_1<i_2<\ldots <i_N}\exp(-\beta\sum_{\ell=1}^N E_{i_\ell})
\end{equation}
or
\begin{equation}\label{15b}
Z_N^+(\beta) = \sum_{i_1\le i_2\le\ldots\le i_N}\exp(-\beta\sum_{\ell=1}^N E_{i_\ell}).
\end{equation}
Let us consider for the moment only systems with a finite
number of energy eigenvalues $E_1,\ldots,E_n$. Expanding the exponentials in
(\ref{15a}),(\ref{15b}) into products and introducing the abbreviations
\begin{equation}\label{16}
x_\ell = \exp(-\beta E_\ell), \quad \ell=1,\ldots,n,
\end{equation}
it is easy to see, c.~f.~section 2, that
\begin{equation}\label{17a}
Z_N^-=f_N(x_1,\ldots,x_n),
\end{equation}
and
\begin{equation}\label{17b}
Z_N^+=b_N(x_1,\ldots,x_n).
\end{equation}
Hence the fermionic/bosonic partitions functions for particular
systems are obtained if the corresponding symmetric polynomials
$f_N$/$b_N$ are evaluated at the values (\ref{16}), where only
$\beta$ is left variable.  The grand canonical
partition functions obviously correspond to the generating
functions of special symmetric polynomials introduced in
(\ref{10}).  It is then obvious that suitable relations between
symmetric polynomials can be translated into relations between
partition functions which are independent of the particular
physical system, e.~g.~of the common potential or the dimension
of the physical space, as long as the systems are ideal gases.

For systems with infinite-dimensional Hilbert spaces,
i.~e.~$n=\infty$, the polynomials degenerate into formal power
series. Evaluation at the values (\ref{16}) leads to
partition functions only if these series converge. But then the
translation of combinatorical results into physical ones is
still possible.\\ Having identified the physical interpretation
of the $f$'s and the $b$'s it remains to investigate the power
sums $s_j$. Since $x_\ell^j = \exp(-j \beta E_\ell)$ we conclude
\begin{equation}\label{18}
s_j(e^{-\beta E_1},\ldots,e^{-\beta E_\ell},\ldots)=Z_1^\pm (j \beta).
\end{equation}
Note that $Z_1^+=Z_1^-$.

Now we can interprete all results mentioned in the last section
as relations between the physical partition functions. We will
not reproduce these relations anew but say a few words on
each. The Newton identities (\ref{7}) are viewed as recursion
relations for partition functions which allow the explicit
calculation of $Z_N^\pm(\beta)$ if $Z_1(\beta)$ is known for all
$\beta$. In this context they appear for the first time in the
textbook of P.~Landsberg \cite{Lan61}. Later they have been
re-discovered and applied recently to systems of few quantum
particles, see
e.g. \cite{BoF:JCP93,BLD:PRE97A,GrH:PRL97,WiW:JMO97,ScS:PA98}. An
explicit representation equivalent to (\ref{11}) appears within
a physical context in \cite{ScS:PA98} and \cite{ScS:PA99}.
Equations (\ref{9}) and (\ref{12}) express relations between
bosonic and fermionic partition functions which have, to our
knowledge, not yet been considered in the physical literature. Note,
however, that the identity (\ref{10b}) can be rewritten as
\begin{equation}\label{19}
{\cal Z}^+(z,\beta) {\cal Z}^-(-z,\beta) =1,
\end{equation}
which obviously follows from the product representations
and is implicitely
contained in the above-mentioned articles based on a ``unified treatment" of Fermi
and Bose statistical mechanics of ideal gases
\cite{Lee:JMP95,Lee:PRE97A,Lee:PRE97B}. We remark that (\ref{19})
implies $Z_1^+=Z_1^-$ which is closely related to the fact
that in the classical limit (i.~e.~$z\rightarrow 0$) the
distinction between Fermi and Bose gases disappears.
Further we note that from the product representation (\ref{10},\ref{10a})
and (\ref{10b})
one immediately derives the well-known fact
that ${\cal Z}^-$ has zeros at
$z_\ell=-\exp(\beta E_\ell)$ and no poles, and hence  ${\cal Z}^+$
has poles at $z_\ell=\exp(\beta E_\ell)$ and no zeros.
Poles and zeros lie outside the physical domain of
$z\in(0,\infty)$ for fermions and $z\in(0,1)$ for bosons.

\section{Concluding remarks}

In this article we have suggested a new way of looking
at the partition functions of ideal quantum gases. The
construction of these functions is split into two steps.
In the first step we consider a finite or infinite set of abstract
energy levels as symbolic variables and define certain symmetric
polynomials (or formal power series) in these variables: the
$b_N$ for bosons and the $f_N$ for fermions. The symmetry involved
is a symmetry between energy levels, not between particles. It
is due to the fact that for non-interacting particles
all possible occupations of the energy levels are counted on an equal footing.
This first step is the
same for each physical system.

The second step consists in an evaluation of the
symmetric polynomials $b_N$ and $f_N$ at the values
$e^{-\beta E_\ell}$ for the variable
$x_\ell$. Only within this step the numerical energy values
of the concrete physical systems, their degeneracies et cetera
enter into the expression for the partition functions. Now the different
possible occupations of the energy levels contribute to the partition function with
different weights.  The mentioned
physically relevant combinatorical results are independent from the second step
and hold for all considered systems.

Finally we would like to mention that the wording ``partition function"
is due to Darwin and Fowler \cite{DF:PM22} and is obviously connected to
the ``partition" of energy into the energy levels of a system. Thus
the combinatorical element was present from the outset, although the
connection to the combinatorical partitions occuring in the explicit
formulae of section 3 seems to be novel.

\section*{Acknowledgement}
We thank Klaus B\"arwinkel, Peter Landsberg, Howard Lee, Marshall Luban, Guy Melançon, and Fabrice Philippe
for stimulating discussions
and/or valuable hints.




\end{document}